\documentclass[prd,preprintnumbers,amsmath,amssymb,floatfix,twocolumn,10pt,nofootinbib]{revtex4-1}
\usepackage{amsmath,amssymb}
\usepackage[usenames,dvipsnames]{xcolor}
\usepackage{graphicx}
\usepackage{epsfig}
\usepackage{rotating}
\usepackage[normalem]{ulem}

\newcommand\bef{\begin{figure}}
\newcommand\eef[1]{\label{fg:#1}\end{figure}}
\newcommand\besf{\begin{subfigure}}
\newcommand\eesf[1]{\label{sfg:#1}\end{subfigure}}
\newcommand\beq{\begin{equation}}
\newcommand\eeq[1]{\label{eq:#1}\end{equation}}
\newcommand\beqa{\begin{eqnarray}}
\newcommand\eeqa[1]{\label{eq:#1}\end{eqnarray}}
\newcommand\bet{\begin{table}}
\newcommand\eet[1]{\label{tb:#1}\end{table}}
\newcommand\best{\begin{subtable}}
\newcommand\eest[1]{\label{stb:#1}\end{subtable}}
\newcommand\betb{\begin{center}\begin{tabular}}
\newcommand\eetb{\end{tabular}\end{center}}
\newcommand\beit{\begin{itemize}}
\newcommand\eeit{\end{itemize}}

\newcommand\fgn[1]{Figure \ref{fg:#1}}
\newcommand\eqn[1]{Eq.\ (\ref{eq:#1})}
\newcommand\scn[1]{Section \ref{sec:#1}}

\newcommand\tbn[1]{Table \ref{tb:#1}}

\newcommand{\nn}{\nonumber}

\newcommand\dadt[4]{[\bar{#1}\bar{#2}]_{3_c}[#3 #4]_{\bar{3}_c}}
\newcommand\dads[4]{[\bar{#1}\bar{#2}]_{\bar{6}_c}[#3 #4]_{6_c}}
\newcommand\dad[4]{[\bar{#1}\bar{#2}]_{\bar{\mathcal{G}}}[#3 #4]_{\mathcal{G}}}

\begin{document}
\title{$X(3872)$ and $Y(4140)$ using diquark-antidiquark operators with lattice QCD  }
\affiliation{Institute of Physics, University of Graz, A-8010 Graz, Austria}
\author{M.\ \surname{Padmanath}}
\email{padmanath.madanagopalan@uni-graz.at}
\author{C. B.\ \surname{Lang}}
\email{christian.lang@uni-graz.at}
\affiliation{Institute of Physics, University of Graz, A-8010 Graz, Austria}
\author{S.\ \surname{Prelovsek}}
\email{sasa.prelovsek@ijs.si}
\affiliation{Department of Physics, University of Ljubljana, Jadranska 19, 1000 Ljubljana, Slovenia}
\affiliation{Jozef Stefan Institute, Jadranska 19, 1000 Ljubljana, Slovenia} 
\affiliation{Theory Center, Jefferson Lab, 12000 Jefferson Avenue, Newport News, Virginia 23606, USA}

\preprint{JLAB-THY-15-2017}

\begin{abstract}
We perform a lattice study of charmonium-like mesons with $J^{PC}=1^{++}$ and  three quark contents 
$\bar cc \bar du$, $\bar cc(\bar uu+\bar dd)$ and  $\bar cc \bar ss$, where the later two  can mix 
with $\bar cc$. This simulation with  $N_f\!=\!2$ and $m_\pi\!\simeq \!266~$MeV  aims at the possible 
signatures of four-quark exotic states. We utilize a  large  basis of $\bar cc$, two-meson 
and diquark-antidiquark interpolating fields, with diquarks in both antitriplet and sextet color 
representations. A lattice candidate for $X(3872)$ with $I\!=\!0$ is observed very close to the 
experimental state only if both $\bar cc$ and $D\bar D^*$ interpolators are included; the candidate is 
not found if diquark-antidiquark and $D\bar D^*$ are used in the absence of $\bar cc$. No 
candidate for neutral or charged $X(3872)$, or any other exotic candidates  are found in the 
$I\!=\!1$ channel. We also do not find signatures of exotic $\bar cc\bar ss$ candidates  below 
$4.2~$GeV, such as $Y(4140)$. Possible physics and methodology related reasons for that are discussed. 
Along the way, we present the  diquark-antidiquark operators as linear combinations of the two-meson 
operators via the Fierz transformations.  
\end{abstract}

\maketitle

\section{Introduction\label{sec:Intro}}
The experimental discovery of  charged resonances  $Z_c(3900)^+$ \cite{Ablikim:2013mio}   and $Z(4430)^{\pm}$ \cite{Choi:2007wga,Mizuk:2009da}  gives signatures
for hadrons with minimal quark content $\bar cc\bar du$.  The neutral $X(3872)$ and yet-unconfirmed $Y(4140)$ with charge parity
$C\!=\!+1$ also appear to have significant four-quark Fock components. Most of the observed  exotic states have 
$J^P=1^+$. The $J^P$ for some  has not been settled experimentally and $J^P=1^+$ presents one possible option.  
      
In this paper, we perform a lattice investigation of the charmonium spectrum, looking for charmonium-like 
states with quantum numbers $J^{PC}=1^{++}$ and three quark contents:  $\bar cc \bar du$, $\bar cc(\bar uu+\bar dd)$ 
and $\bar cc \bar ss$, where the later two channels have $I\!=\!0$ and can mix with $\bar cc$ ($C$  indicates $C$-parity of neutral isospin partners for charged states). Our main interest 
in these channels is aimed at a first-principle  study of  $X(3872)$ and $Y(4140)$, which were observed in 
$X(3872)\rightarrow J/\psi \rho,~J/\psi\, \omega,~ D\bar{D}^*$ and $Y(4140)\rightarrow J/\psi \phi$, for example. 

From the experimental side, the long known exotic candidate $X(3872)$ \cite{Choi:2003ue}  
is confirmed to have $J^{PC}=1^{++}$ \cite{Aaij:2013zoa}. 
However, questions about its isospin remain unsettled. If it has  
isospin $I=1$, one expects charged partners. Observation of a nearly equal branching fraction for 
$X(3872) \rightarrow J/\psi\, \omega$ and $X(3872) \rightarrow J/\psi \,\rho$ decays \cite{Agashe:2014kda} and searches for 
charged partner $X(3872)$ states decaying to $J/\psi \rho^{\pm}$ \cite{Aubert:2004zr} speak against 
a pure $I=1$ state. There are a few other candidates with $C\!=\!+1$ that could possibly have 
$J^{PC}=1^{++}$ like $X(3940)$ \cite{Abe:2004zs}, $Z(4050)^{\pm}$ \cite{Mizuk:2009da} and 
$Z(4250)^{\pm}$ \cite{Mizuk:2009da}. A detailed review on these can be found in Ref. \cite{Olsen:2014qna}.

The growing evidence for the $Y(4140)$ resonance in the $J/\psi \phi$ invariant mass 
\cite{Aaltonen:2009tz} serves as promising signature for exotic hadrons with 
hidden strangeness. Similarities in the properties of $X(3930)$ and $Y(4140)$ led to an interpretation that 
$X(3930)$ may be a $D^*\bar D^*$ molecule and $Y(4140)$ is its hidden strange counterpart $D_s^*D_s^*$ molecule \cite{Liu:2010hf}. 
However, the upper limit for the production of $Y(4140)$ in $\gamma\gamma\rightarrow  J/\psi\,\phi$ is observed to be much lower 
than theoretical expectations for a $D_s^*D_s^*$ molecule with $J^{PC}=0^{++}$ and $2^{++}$\cite{Shen:2009vs}.
Hence the quantum numbers of $Y(4140)$ stay unsettled and  it remains open for a $J^{PC}=1^{++}$ assignment.

From a theoretical perspective, the description of such resonances is not settled. Several 
suggestions have been made interpreting them as mesonic molecules \cite{Swanson:2003tb}, as
diquark-antidiquark structures \cite{Maiani:2004vq}, as a cusp phenomena \cite{Bugg:2004sh} or as 
a $|c\bar{c}g\rangle$ hybrid meson \cite{Close:2003sg}. A great deal of theoretical studies are based on 
phenomenological approaches like quark model, (unitarized) effective field theory 
and QCD sum rules (see reviews  \cite{Olsen:2014qna}).
  
It is paramount to establish whether QCD supports the existence of resonances with exotic character using first principles 
techniques such as lattice QCD. Simulations that considered only $\bar{c}c$ interpolators could not provide evidence for 
$X(3872)$. The first evidence from a lattice simulation for $X(3872)$ with $I\!=\!0$ was reported in Ref. \cite{Prelovsek:2013cra}, where 
a combination of $\bar cc$ as well as  $D\bar D^*$ and $J/\psi \omega$  interpolators was used. Recently, another calculation using the Highly Improved Staggered Quark action also gave evidence for $X(3872)$, using $\bar{c}c$ and $D\bar{D}^*$ interpolating fields 
\cite{Lee:2014uta}. The search for the $Y(4140)$ resonance was performed only in  \cite{Ozaki:2012ce}, where  a phase shift for $J/\psi \phi$ scattering in $s$-wave and $p$-wave was extracted from $N_f\!=\!2+1$ simulation using twisted boundary conditions, and neglecting strange-quark annihilation. The resulting phase shifts did not support existence of a resonance. 

The novel feature of the present study is to add diquark-antidiquark $\dad{c}{q}{c}{q}$ operators to the basis of interpolating fields and to extend the
extraction of the charmonium spectrum with $J^{PC}=1^{++}$ to a higher energy range.
This is the first dynamical lattice calculation involving diquark-antidiquark operators 
along with several two-meson and $\bar cc$ kind of interpolators to study $X(3872)$ and $Y(4140)$.  
We consider the color structures ${\mathcal G}=\bar 3_c,6_c$  for diquarks, which
have been suggested already in the late seventies \cite{Jaffe:1976ih}. Recently many phenomenological studies 
\cite{Maiani:2004uc,Maiani:2004vq} and a few lattice studies \cite{Chiu:2006hd,Chiu:2005ey}
used them  to extract the light and heavy meson spectra. 
In Ref. \cite{Chiu:2006hd} a calculation using two-meson and diquark-antidiquark interpolators was performed to investigate 
mass spectrum of $1^{++}$ exotic mesons in quenched lattice QCD. However,  only one energy level was extracted, which is not sufficient to provide evidence for $X(3872)$ or $Y(4140)$.  

In this paper we address the following questions: Is the lattice candidate for $X(3872)$ reproduced in presence of 
diquark-antidiquark operators? Which are the crucial operator structures for its emergence? How important are the 
$\dad{c}{q}{c}{q}$ Fock components in the established $X(3872)$?  Do we find a lattice candidate for charged or 
neutral $X(3872)$ with $I\!=\!1$? Do operators with hidden strangeness render a candidate for $Y(4140)$? Do we 
find candidate states for other possible exotic states in the channels being probed?  
 
The paper is organized as follows. \scn{Interpolators} addresses the expected two-meson scattering channels below 4.2 GeV. The lattice methodology is discussed in Sect. \ref{sec:methodology}. In Sect. \ref{sec:Fierz} and the Appendix we discuss the relations 
between our diquark-antidiquark and two-meson interpolators via  Fierz transformations. \scn{Results} is dedicated to results 
and we conclude in Sect. \ref{sec:Conc}.

\section{Two particle states in lattice QCD\label{sec:Interpolators}}

A major hurdle in excited-state spectroscopy is that most of the states lie above various 
thresholds and decay strongly in experiments. All states carrying the same quantum numbers, 
including the single-particle and multiparticle states, in principle contribute to the 
eigenstates of the Hamiltonian. The determination of scattering properties relies 
on precise identification of all the eigenstates below and close above the energy 
of our interest. The continuous spectrum of scattering states in the continuum gets 
reduced to a discrete set of eigenstates, because lattice momenta are discretized due to 
the finite lattice size.  

Considering two-meson states with total momentum zero and without interaction, their energies are just the sum of the
individual particle energies 
\begin{equation}
\label{eni}
E^{n.i.}_{M_1(\mathbf{n})M_2(-\mathbf{n})}= E_1(p) + E_2(p),\ p=\frac{2\pi |\mathbf{n}|}{L},\  \mathbf{n}\in N^3. 
\end{equation}
In the presence of interactions, the energies get shifted depending on the interaction strength.  For our lattice setting the noninteracting two-meson levels 
with  $J^{PC}=1^{++}$, total momentum zero  in the indicated energy range are
\beit

\item ~~$I = 0$; ~$\bar cc(\bar uu+\bar dd)$ and $\bar cc$; ~$E\lesssim4.2~$GeV 
\begin{align*}
& D(0)\bar D^*(0), & & J/\psi(0)\omega(0), & & D(1)\bar D^*(-1),\\ 
 &J/\psi(1)\omega(-1), & &\eta_c(1)\sigma(-1), & & \chi_{c1}(0)\sigma(0)\;. 
\end{align*}
\item ~~$I = 1$; ~$\bar cc \bar du$ ; ~$E\lesssim4.2~$GeV
\begin{align*}
&D(0)\bar D^*(0),& &J/\psi(0)\rho(0),& &D(1)\bar D^*(-1),\\
&J/\psi(1)\rho(-1),& &\chi_{c1}(1)\pi(-1),& &\chi_{c0}(1)\pi(-1)\;. 
\end{align*}
\item ~~$I = 0$; ~$\bar cc \bar ss$  and $\bar cc$; ~$E\lesssim4.3~$GeV
\begin{align*}
&D_s(0)\bar D_s^*(0),& &J/\psi(0)\phi(0),& &D_s(1)\bar D_s^*(-1),\\
&J/\psi(1)\phi(-1)\;.&&&&
\end{align*}
\eeit
The parentheses denote meson momenta in units of $2\pi/L$. 

We consider the flavor sectors $\bar cc(\bar uu+\bar dd)$ and $\bar cc \bar ss$ 
separately. In nature these two $I=0$ sectors  can mix and they could in principle mix 
also in our simulation without dynamical strange quarks. However, if both flavor 
sectors would be treated together, then  $6+4=10$ two-particle $I=0$ states are expected 
below $4.2~$ GeV. This would make the resulting  spectrum denser and noisier, so  the 
identification of eigenstates and the search for exotics would be even more challenging. 
We therefore consider these two sectors separately in this first search for possible 
exotics in the   extended energy region. The corresponding  assumptions will be 
discussed for each flavor channel along with the results.  
 
The noninteracting energies will be shown by the horizontal lines in our plots, and 
follow from the masses and the single meson energies determined on the same set of gauge 
configurations \cite{Mohler:2012na,Lang:2011mn,Lang:2014tia}. The energies of the $\sigma$ meson 
 using single-hadron approximation are $a\,m_\sigma= 0.302(15)$ and $a\,E_{\sigma(1)}=0.534(22)$.
Including two-meson operators up to 4.2 GeV at $m_{\pi}=266$~MeV should be sufficient 
in searching for narrow exotic candidates below 4.2 GeV. Details of all the interpolators 
used, including the diquark-antidiquark  interpolators, can be found in the next section. 

The mesons $R=\rho,\sigma$ are resonances that decay to  $\pi\pi$ or $\pi\eta$ in QCD with $N_f\!=\!2$. 
A proper simulation which would consider the three-meson system  \cite{Polejaeva:2012ut}  has not been performed 
in practice yet. In absence of this, a simplifying approximation for channels containing these resonances is 
adopted. We determine the energy of $R(p)$ as the ground state energy obtained from the correlation matrix 
with $\sum_x e^{ipx} \bar q(x)\Gamma q(x)$ interpolators. This energy is used for the horizontal lines in 
the plots. This basis renders in all cases just one low-lying state. Within our approximation this low-lying 
state corresponds to a resonance $R$ with momentum $p$, to a two-particle state $\pi\pi/\pi\eta$ with total 
momentum $p$, or to some mixture of $R$ and the two-particle state. We also do 
not consider nonresonant three-meson levels which could appear above  $\eta_c \pi\pi$, $J/\psi \pi\pi$, $\eta_c K\bar K$, $J/\psi K\bar K$ thresholds. 
Based on the experience with two-meson operators we do not expect that without 
explicit incorporation of three-meson interpolating fields these three-meson states appear in the spectra.

\section{Lattice methodology\label{sec:methodology}}

These calculations are performed on $N_f\!=\!2$ dynamical gauge configurations with 
$m_{\pi}\!\simeq \!266$ MeV \cite{Hasenfratz:2008ce} and  with other parameters provided 
in \tbn{latpar}. The mass-degenerate $u/d$ quarks are  based on a tree-level improved 
Wilson-clover action. The strange quark is present only in the valence sector and we 
assume that the valence strange content could uncover hints on the possible existence of 
the $\bar cc\bar ss$ exotics.  The absence of dynamical strange quarks prevents 
$\bar cc\bar ss$ intermediate states  in the  $\bar cc(\bar uu+\bar dd)$ and $\bar cc$ 
sector, in accordance with treating these  two $I=0$ sectors separately in our study. 
With a rather small box size of $L\!\simeq\! 2$ fm, one expects to have large finite 
size effects. On the other hand this serves as a crucial practical advantage by reducing 
the number of two-meson scattering states $M_1(\mathbf{n})M_2(-\mathbf{n})$ in the 
energy range of our interest. This helps in easier identification of the possible 
resonances that could exist along with the regular two-meson energy levels. It also 
reduces computational cost as one needs to consider a smaller number of distillation 
eigenvectors and two-meson interpolators with respect to a study in larger volume.

\bet[tbh]
\centering
\betb{ccccccccccccc}
\hline
Lattice size     & $\kappa$ & $\beta$ &  $N_{\mathrm{cfgs}}$ &  $m_\pi$ [MeV] &    $a$ [fm]    & $L$ [fm]  \\\hline
$16^3 \times 32$ & $0.1283$ &   7.1   &        280           &    266(3)(3) & $0.1239(13)$ &  1.98     
\\\hline
\eetb
\caption{Details of the gauge field ensemble used.}
\eet{latpar} 

We construct altogether 22 interpolators with $J^{PC}=1^{++}$ and total momentum zero for the three cases of our 
interest ($T_1^{++}$ irreducible representation of the discrete lattice group $O_h$ is employed):
\begin{widetext}
\beqa
O_{1-8}^{\bar cc}&=& \bar c \hat{M} c(0) ~\tfrac{1}{2}(1+K_d), \qquad\mbox{see Table X of Ref. \cite{Mohler:2012na}}\  \label{operators0} \\ 
O_9^{MM}&=&\bar c \gamma_5 u(0)~\bar u\gamma_i c(0) - \bar c \gamma_i u(0)~\bar u\gamma_5 c(0) +K_d\{u\to d\}, \  \nonumber \\
O_{10}^{MM}&=&\epsilon_{ijk} ~\bar c \gamma_j c(0)~\{~\bar u\gamma_k u(0)+K_d\{u\to d\}~\},  \nonumber \\ 
O_{11}^{MM}&=&\!\!\!\!\sum_{e_p=\pm e_{x,y,z}}\{\bar c \gamma_5 u(e_p)~\bar u\gamma_i c(-e_p) - \bar c \gamma_i u(e_p)~\bar u\gamma_5 c(-e_p)\} +K_d\ \{u\to d\}, \nonumber\\ 
O_{12}^{MM}&=&\bar c \gamma_5 \gamma_4 u(0)~\bar u\gamma_i \gamma_4 c(0) -\bar c \gamma_i \gamma_4 u(0)~\bar u\gamma_5 \gamma_4 c(0)+K_d\{u\to d\}, \nonumber\\ 
O_{13}^{MM}&=&\epsilon_{ijk} ~\bar c \gamma_j \gamma_4 c(0)~\{\bar u\gamma_k \gamma_4 u(0)+K_d \{u\to d\}\}, \nonumber\\ 
O_{14}^{MM}&=&\sum_{e_p=\pm e_{x,y,z}}\epsilon_{ijl} ~\bar c \gamma_j c(e_p)~\{\bar u\gamma_l u(-e_p)+K_d \{u\to d\}\},\nonumber\\
O_{15}^{MM}&=&\{\bar c\gamma_5 c (e_p) ~\bar uu(-e_p)~-~\bar c\gamma_5 c (-e_p) ~\bar uu(e_p)\}_{p=i}+K_d \{u\to d\}, \nonumber\\ 
O_{16}^{MM}&=&\epsilon_{ijp} \{\bar c\gamma_j\gamma_5 c (-e_p) ~\bar u\gamma_5u(e_p)~-\bar c\gamma_j\gamma_5 c (e_p) ~\bar u\gamma_5u(-e_p)\} +K_d \{u\to d\},\nonumber\\ 
O_{17}^{MM}&=&\bar c\gamma_i\gamma_5 c (0) ~\bar uu(0) +K_d \{u\to d\},\nonumber\\ 
O_{18}^{MM}&=&\{\bar c c (e_p) ~\bar u\gamma_5u(-e_p)~-~\bar cc (-e_p) ~\bar u\gamma_5 u(e_p)\}_{p=i} +K_d \{u\to d\} ,\nonumber\\ 
O_{19}^{4q}&=&[ \bar c~ C\gamma_5 \bar u^T]_{3_c}[ c^T \gamma_i C u]_{\bar 3_c} +  [ \bar c~ C\gamma_i \bar u^T]_{3_c}[ c^T \gamma_5 C u]_{\bar 3_c}+K_d \{u\to d\}, \nonumber\\
O_{20}^{4q}&=&[ \bar c~ C \bar u^T]_{3_c}[ c^T \gamma_i \gamma_5 C u]_{\bar 3_c} +  [ \bar c~ C \gamma_i \gamma_5 \bar u^T]_{3_c}[ c^T C u]_{\bar 3_c}+K_d \{u\to d\}, \nonumber\\ 
O_{21}^{4q}&=&[ \bar c~ C\gamma_5 \bar u^T]_{\bar 6_c}[ c^T \gamma_i C u]_{6_c} +  [ \bar c~ C\gamma_i \bar u^T]_{\bar 6_c}[ c^T \gamma_5 C u]_{6_c}+K_d \{u\to d\}, \nonumber\\
O_{22}^{4q}&=&[ \bar c~ C \bar u^T]_{\bar 6_c}[ c^T \gamma_i \gamma_5 C u]_{6_c} +  [ \bar c~ C \gamma_i \gamma_5 \bar u^T]_{\bar 6_c}[ c^T C u]_{6_c}+K_d \{u\to d\}. \nn 
\eeqa{operators1}
\end{widetext}   
The indices $i, j, k$ and $l$ define the Euclidean Dirac gamma matrices, while the index $p$ indicates the momentum 
direction. Einstein's summation convention is implied for repeated indices. The unsummed index $i$ in all the operators 
defines the polarization. The $C=i\gamma_2\gamma_4$ is the charge conjugation matrix. The coefficient $K_d$ depends on the 
quark content:  $K_d\!=\!1$ is used for  $\bar cc(\bar uu+\bar dd)$ and  $K_d\!=\!0$ 
for $\bar cc \bar ss$ followed by using strange quark propagators instead of the light quark propagators. For $I=1$ channel 
we apply $K_d\!=\!-1$ which gives the flavor content $\bar cc(\bar uu-\bar dd)$ and has the same spectrum as $\bar cc\bar du$ 
in the isospin limit.

We emphasize the use of four operators $O^{4q}$ with diquark-antidiquark structure and color antitriplet or sextet diquarks
\beqa
[ \bar c \Gamma_1 \bar q]_{\mathcal{G}}[ c \Gamma_2 q]_{\mathcal{\bar G}} &\equiv& \sum_{\mathbf{x_1}} \mathcal{G}_{ab_1c_1}\bar c_{b_1}^{\alpha_1}\Gamma_1^{\alpha_1\beta_1}\bar q^{\beta_1}_{c_1}(\mathbf{x_1},t_f) \nonumber \\
 ~&\cdot& \sum_{\mathbf{x_2}}\mathcal{G}_{ab_2c_2}c_{b_2}^{\alpha_2}\Gamma_2^{\alpha_2\beta_2}q^{\beta_2}_{c_2}(\mathbf{x_2},t_f).
\eeqa{4q}
Here  $a=1,2,3$ for color triplet and $a=1,...,6$ for sextet, while $b,c=1,2,3$ for both:
\beqa
\mathcal{G}^{3}_{abc} = \mathcal{G}^{\bar 3}_{abc} &=&\epsilon_{abc}~  \\
\mathcal{G}^{6}_{abc}=\mathcal{G}^{\bar 6}_{abc}&=&1 ~\mbox{:}~ a=1,2,3 ~\mbox{and}~ a\ne b\ne c \nn \\
\mathcal{G}^{6}_{abc}=\mathcal{G}^{\bar 6}_{abc}&=&\sqrt{2}~\mbox{:}~ a=4,5,6 ~\mbox{and}~ a-3= b = c \nn
\eeqa{4qCG}
while the remaining $\mathcal{G}_{abc}$ are zero. The operator [\eqn{4q}]  reduces to   
$\sum_{\mathbf{x}} \bar c(\mathbf{x}) \bar q(\mathbf{x}) c(\mathbf{x}) q(\mathbf{x})$ on ensemble averaging,
where the gauge configurations are not gauge fixed.

The interpolators are related with the two-meson channels as listed in \tbn{operators2}. noninteracting levels corresponding 
to some of these two-meson channels lie above our energy of interest, and the corresponding 
interpolators are not considered.

\bet
\betb{c | c | c | c }
\hline
N    & $\bar cc(\bar uu+\bar dd)$   & $\bar cc\bar ud$      & $\bar cc\bar ss$ \\\hline\hline
$O_{1-8}^{\bar cc}$   &    $\bar c ~\hat{M}~c$   & Does not couple        &    $\bar c ~\hat{M}~c$     \\\hline
$O_{9}^{MM}$          &    $D(0)\bar D^*(0)$     & $D(0)\bar D^*(0)$      &  $D_s(0)\bar D_s^*(0)$   \\\hline
$O_{10}^{MM}$         & $J/\psi(0)\omega(0)$     & $J/\psi(0)\rho(0)$     & $J/\psi(0)\phi(0)$  \\\hline
$O_{11}^{MM}$         &    $D(1)\bar D^*(-1)$    & $D(1)\bar D^*(-1)$     &  $D_s(1)\bar D_s^*(-1)$   \\\hline
$O_{12}^{MM}$         &    $D(0)\bar D^*(0)$     & $D(0)\bar D^*(0)$      &  $D_s(0)\bar D_s^*(0)$   \\\hline
$O_{13}^{MM}$         & $J/\psi(0)\omega(0)$     & $J/\psi(0)\rho(0)$     & $J/\psi(0)\phi(0)$  \\\hline
$O_{14}^{MM}$         & $J/\psi(1)\omega(-1)$    & $J/\psi(1)\rho(-1)$    & $J/\psi(1)\phi(-1)$ \\\hline
$O_{15}^{MM}$         & $\eta_c(1)\sigma(-1)$    & $\eta_c(1)a_0(-1)$     &  Not used            \\\hline
$O_{16}^{MM}$         & $\chi_{c1}(1)\eta(-1)$   & $\chi_{c1}(1)\pi(-1)$  &  Not used            \\\hline
$O_{17}^{MM}$         & $\chi_{c1}(0)\sigma(0)$  & $\chi_{c1}(0)a_0(0)$   &  Not used            \\\hline
$O_{18}^{MM}$         & $\chi_{c0}(1)\eta(-1)$   & $\chi_{c0}(1)\pi(-1)$  &  Not used            \\\hline
$O_{19-20}^{4q}$      & $\dadt{c}{q}{c}{q}$      & $\dadt{c}{u}{c}{d}$    &  $\dadt{c}{s}{c}{s}$  \\\hline
$O_{21-22}^{4q}$      & $\dads{c}{q}{c}{q}$      & $\dads{c}{u}{c}{d}$    &  $\dads{c}{s}{c}{s}$  \\\hline
\eetb
\caption{List of interpolators ($J^{PC}=1^{++}$) and their correspondence with various two-meson 
scattering channels.}
\eet{operators2}

The Wick contractions considered in the computation of the correlation functions are shown in \fgn{Wickcont}. There are two 
other classes of diagrams, which are not considered: one in which no valence quark propagates from source to sink, and the 
other class in which only the light/strange quarks propagate from source to sink and the $\bar{c} c$ pair annihilates. The 
effects from these two classes of diagrams, with the charm quark not propagating from source to the sink, are known to be suppressed 
due to the Okubo-Zweig-Iizuka rule. They correspond to mixing with a number of channels  that contain only the $u/d$ and 
$s$ quarks, which represents currently unsolved challenge in lattice QCD.  Note that the annihilation of $u/d$ and $s$ quarks 
as well as  mixing with $\bar cc$ is taken into account, unlike in the simulation \cite{Ozaki:2012ce} aimed at $Y(4140)$, for example.

\begin{widetext}

\begin{figure*}[tbh]
\centering
\hspace{-1.5cm}
\parbox{.45\linewidth}{
\centering
  \includegraphics[scale=0.35]{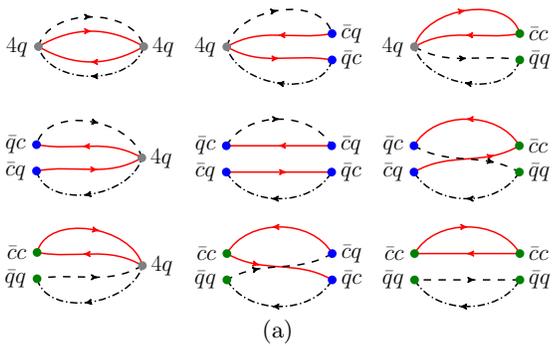}\\
(a)}\hspace{0.25cm}
\hspace{0.25cm}
\parbox{.45\linewidth}{ 
\centering
  \includegraphics[scale=0.3]{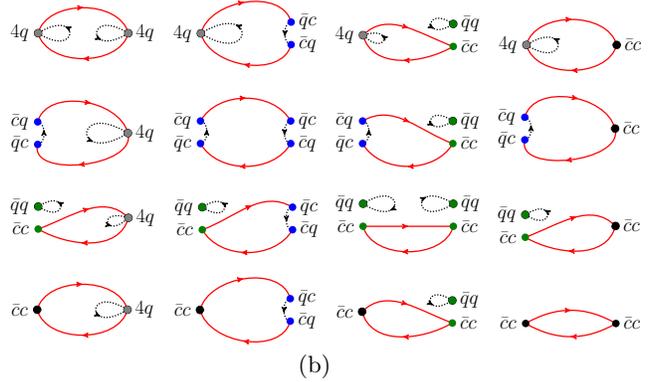}\\
(b)}
\caption{The Wick contractions considered in our calculations. (a) Connected contraction diagrams. (b) Diagrams, 
in which the light/strange quarks do not propagate from source to sink. The correlation functions in the 
$\bar cc(\bar uu+\bar dd)$ and $\bar cc\bar ss$ cases are linear combinations of the diagrams of kind (a) and (b), 
while the correlation functions between the operators with quark content $\bar cc\bar ud$ are constructed purely 
from diagrams of kind (a).}
\label{fg:Wickcont}
\end{figure*}

\end{widetext}

Using the interpolators  listed in Eq. (\ref{operators0}) and \tbn{operators2}, we compute the full coupled
correlation functions 
\beq
\mathcal C_{jk}(t)=\langle\Omega|O_j (t_{s}+t) O_k^\dagger (t_{s})|\Omega\rangle =\sum_{n}Z_k^{n*}Z_j^ne^{-E_n t}. 
\eeq{Corr}
For an efficient computation of these correlation matrices, we utilize the ``distillation" method for 
the quark sources as proposed in Ref. \cite{Peardon:2009gh}. In this method the quark sources are 
build from the $N_v$ lowest eigenmodes of the gauge-covariant Laplacian on a given time slice, $t_{s}$.
We use $N_{v}\!=\!64$ for computation of correlators involving $u/d$ quarks, while for the correlators 
with hidden strange content, we use $N_{v}\!=\!48$. The correlation functions with $u/d$ quarks are
computed only for polarization along the $x$-axis and averaged over all $t_{s}$, while correlation functions 
involving hidden strange quarks are averaged over all polarizations and for all even values of $t_{s}$.

The energies $E_n$ and overlaps $Z_j^{(n)}=\langle \Omega|O_i|n\rangle$  for all eigenstates $n$ are extracted using 
the well-established generalized eigenvalue problem \cite{Michael:1985ne} 

\beq
\mathcal C(t)\,u^{(n)}(t) = \lambda^{(n)}(t,t_0)\,\mathcal C(t_0)\,u^{(n)}(t).
\eeq{gevp1} 
The energies $E_n$ are extracted asymptotically from two-exponential fits to the eigenvalues 

\beq \lambda^{(n)}(t,t_0)\propto A_n e^{-E_n t} + A'_n e^{-E'_n t}~,\quad E_n^\prime>E_n. \eeq{texpf}
We find consistent results for $t_0=2, 3$ and present the results for $t_0=2$. 
The two-exponential fits were typically done in the range $3\leq t \leq 14$.
The eigenvectors $u^{(n)}$ determine the overlaps 

\beq
Z_j^{(n)}(t)=e^{E_nt/2}\frac{|\mathcal C_{jk}(t)u_k^{(n)}(t)|}{|\mathcal C(t)u^{(n)}(t)|}\;.
\eeq{gevp2}
The statistical errors obtained using single-elimination jackknife analysis are quoted throughout. 

The complete basis was used in the initial analysis, which was later reduced to an optimized basis, 
separately in each of the three cases, based on a systematic operator pruning. This procedure is 
aimed at getting better signals (in terms of the numbers of states and the quality of the effective 
mass plateau and the overlap factors) in comparison with the spectrum extracted from the full set of operators. After 
finalizing the optimized set of two-meson interpolators, we fixed the $\bar{c}c$ and $\dad{c}{q}{c}{q}$ 
operators that give good signals for a maximum number of extractable states below 
4.2 GeV. The optimized basis that we used for the three cases of quark content are
\beqa
   \bar cc(\bar uu+\bar dd) :     & \quad &O_{1,3,5}^{\bar cc},~O_{9-12,14,15,17}^{MM},~O_{19,21}^{4q}  \nonumber \\ 
   \bar cc\bar ud :& \quad &O_{9-16,18}^{MM},~O_{19,21}^{4q}   \nonumber \\ 
   \bar cc\bar ss :& \quad &O_{1,5}^{\bar cc},~O_{9-11,14}^{MM},~O_{19,21}^{4q}\;.   
\eeqa{optmbasis}

Our principal aim is to find out whether QCD supports exotic states in addition to the conventional charmonia and 
the two-meson scattering levels, which inevitably appear in dynamical QCD. Analytic techniques 
have been proposed for the determination of the scattering matrix for coupled two-hadron scattering channels based on 
L\"uscher-type finite volume formalisms \cite{Luscher:1990ux}. 
These would in principle allow extraction of the masses and decay widths for resonances of interest. A number of 
lattice calculations have already dealt with resonances and shallow bound states in the elastic scattering (see 
\cite{Dudek:2012xn} and \cite{Mohler:2013rwa} for an example of each). The first calculation of a scattering matrix 
for two coupled channels also promises progress in this direction \cite{Dudek:2014qha}. However, such an analysis 
is beyond the scope of current lattice simulations for more than two coupled channels and/or three-hadron scattering
channels, which applies to the case considered.  

Therefore we take a simplified approach, where the existence of possible exotic states is investigated by analyzing 
the number of energy levels, their positions and overlaps with the considered lattice operators 
$\langle \Omega |O_j|n\rangle$. 
 The formalism does predict an appearance of a level in addition to the (shifted) two-particle levels if there is a 
relatively narrow resonance in one channel. We have, for example, found additional levels related to the resonances 
$\rho$ \cite{Lang:2011mn}, $K^*(892)$ \cite{Prelovsek:2013ela}, $D^*_0(2400)$ \cite{Mohler:2012na}, and the bound 
state $D_{s0}^*(2317)$ \cite{Mohler:2013rwa}. Additional levels related to $K_0^*(1430)$ \cite{Dudek:2014qha} and 
$X(3872)$ \cite{Prelovsek:2013cra} have been found in the simulations of two coupled channels. Based on this experience, 
we expect an additional energy level if an exotic state is of similar origin, i.e. if it corresponds to a pole of 
the scattering matrix near the physical axis.  
   
Consider a noninteracting situation. Several two-meson operators considered in \tbn{operators2} contain the vector 
meson $V(1)$ with one unit of momentum. This can reside in irreducible representations (irreps) $A_1$ or $E_2$ of 
the corresponding symmetry group $Dic_4$  \cite{Moore:2006ng,Thomas:2011rh}.  One expects two degenerate energy 
levels for $P(1)V(-1)$ since there are two ways to combine  the vector-meson irrep ($A_1$, $E_2$) with the 
pseudoscalar-meson irrep ($A_2$) to obtain the rest frame irrep of interest $T_1^+$  (see Table III of 
\cite{Moore:2006ng}). The underlying reason is that $PV$ state with $J^P=1^+$ can be in  $s$-wave or in 
$d$-wave (also in continuum) \cite{Briceno:2014oea,Briceno:2013lba,Briceno:2013bda}. In the limit of small coupling 
between $s-$ and $d-$wave, one energy level is due solely to the $s$-wave and the other one to $d$-wave 
\cite{Briceno:2013lba,Briceno:2013bda}.\footnote{The $PV$ spectrum resembles (in the noninteracting limit) 
the spectrum in the deuterium  channel $pn$,  since $S=1$,  $J^P=1^+$ and $l=0,2$ apply in both cases. Figure 2 
of \cite{Briceno:2013lba} indicates that one level $n(1)p(-1)$ is related mostly to $s$-wave and the other to $d$-wave.
L\"uscher's quantisation condition \cite{Briceno:2014oea} does not depend on the spins of the individual particles, 
but on their total spin $S$. } We implement only the $s$-wave interpolator $O^{P(1)V(-1)}$ [Eq. (\ref{operators0})] 
and therefore expect to see only one energy level; this is verified in our observed spectra shown in Sect. 
\ref{sec:Results}.  One would need to employ two distinct interpolators    in order to find two $P(1)V(-1)$ energy 
levels, but the extraction of such  eigenstates   has not been attempted yet for two-meson systems in QCD to our 
knowledge. Our two-meson operators contain also  $V_1(1)V_2(-1)$, where three  levels are expected based on analogous 
arguments \cite{Moore:2006ng}; we expect to find only one level related to $s$-wave interpolators  
[Eq. (\ref{operators0})], and indeed  we do not find two other levels related to $d$-wave  (for total spins 
$S\!=\!1,2$). We emphasize that the omission of additional interpolator structures and avoidance of levels related 
to $d$-waves makes the  search for possible exotics within our approach less cumbersome and results more transparent.
\footnote{If one would find an extra state near $V(1)P(-1)$ or $V_1(1)V_2(-1)$, one would indeed have to identify 
whether this extra state  arises due to the presence of the $d$-wave or  is related to exotics. We do not  address 
this question since we do not find such an extra state.}

Charm quarks being heavy are subject to large discretization errors. We treat the charm quarks using the Fermilab 
formulation \cite{ElKhadra:1996mp}, according to which we tune the charm quark mass by equating the spin averaged 
kinetic mass of the $1S$ charmonium to its physical value. With this formulation, the discretization errors are 
highly suppressed in the energy splitting $E_n-m_{s.a.}$, $m_{s.a.} = \frac14(m_{\eta_c}+3m_{J/\psi})$, which will 
be compared with the experiments. We utilized this method in our earlier calculations on this ensemble and found good 
agreement with the experiments for conventional charmonium in Ref. \cite{Mohler:2012na} as well as, for masses and 
widths of charmed mesons in Refs. \cite{Mohler:2012na,Lang:2014yfa,Mohler:2013rwa}.


\section{Fierz relations \label{sec:Fierz} }
 
The diquark-antidiquark operators  $\dadt{c}{q}{c}{q}$ and $\dads{c}{q}{c}{q}$  can be expressed as linear 
combinations of color singlet currents $(\bar cc)_{1_c}(\bar qq)_{1_c}$ and $(\bar cq)_{1_c}(\bar qc)_{1_c}$ 
\cite{Maiani:2004vq,Ali:2014dva}. These relations are obtained for local currents via Fierz rearrangement 
\cite{Nieves:2003in} and are presented in the Appendix. Note that our quarks are smeared and each meson in $O^{MM}$ 
has definite momentum, but the Fierz relation suggests that $O^{4q}$ and $O^{MM}$ are still linearly dependent. 
 
The Fierz rearrangement is the key idea behind Coleman's argument \cite{Coleman} that in the large $N_c$ limit
application of Fermion quadrilinears to the vacuum creates meson pairs and nothing else. In the physical world 
with $N_c\!=\!3$, it is argued that tetraquarks could exist at subleading orders \cite{Weinberg:2013cfa} 
of large $N_c$ QCD. However, in the presence of the leading order two-meson terms, one should 
take caution in interpreting the nature of the levels purely based on their overlap factors onto various 
four-quark interpolators.

\begin{figure}[tb]
\centering
\includegraphics[scale=0.33]{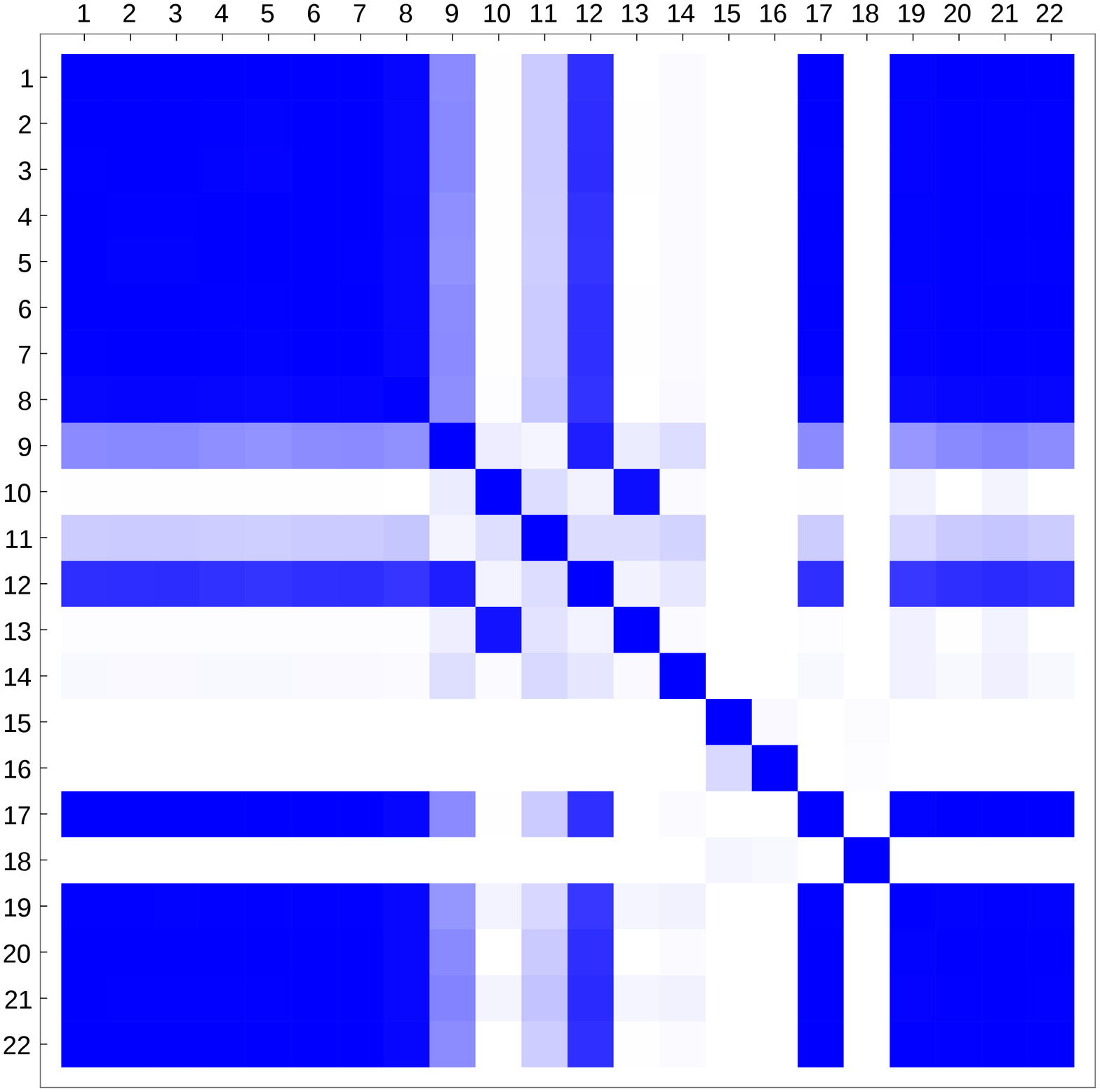}
\includegraphics[scale=0.07]{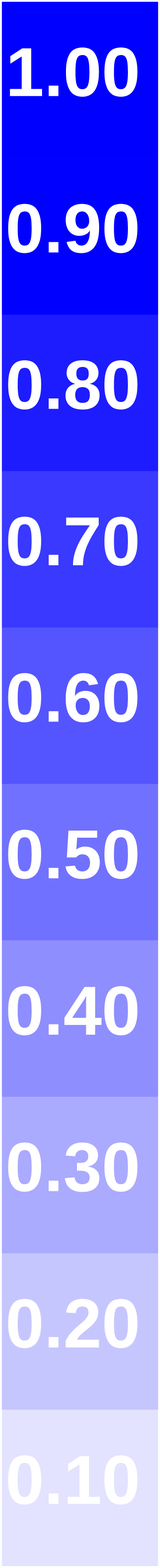}\\
\caption{Time averaged normalized correlation matrix $\tilde{\mathcal C}$ [\eqn{TANCM}] for the operator basis 
$O_{1-22}$ with quark content $\bar cc(\bar uu+\bar dd)$ and $\bar cc$. The axis ticks correspond to the order of 
operators used in Eq. (\ref{operators0}).}
\label{fg:cormat}
\end{figure}

Let us consider a comparative study between the lattice correlators  and the Fierz expansion of $O^{4q}$ operators. 
From Eq. (\ref{opfq1}), we see that the first and second terms in the Fierz expansion represent $D\bar D^*$, while the 
seventh term is similar to the $O^{MM}_{17}=\chi_{c1} ~ \sigma$. Hence we expect significant correlations between 
these operators. This is indeed verified in \fgn{cormat}, showing the time averaged normalized ensemble averaged 
correlation matrix  
\beq
\tilde{\mathcal{C}}_{ij} = \frac19\sum_{t=2}^{10} \frac{\bar{\mathcal C}_{ij}(t)}{\sqrt{\bar{\mathcal C}_{ii}(t)\bar{\mathcal C}_{jj}(t)}}.
\eeq{TANCM}
With this normalization all the diagonal entries are forced to unity and all the off-diagonal entries to be less than unity. 
The $\dad{c}{q}{c}{q}=O^{4q}_{19-22}$ have large correlations onto the $D\bar D^*=O^{MM}_{9,11,12}$ and 
$\chi_{c1} ~ \sigma=O^{MM}_{17}$. The strong correlations between $O^{4q}$ and $O^{cc}$ operators can also be 
explained by the  $\chi_{c1} \sigma$ component in $O^{4q}$, where $\sigma$ couples to the vacuum. 

\begin{figure*}[tbh]
\hspace{-1.0cm}
\parbox{.45\linewidth}{
\centering
\includegraphics[scale=0.72]{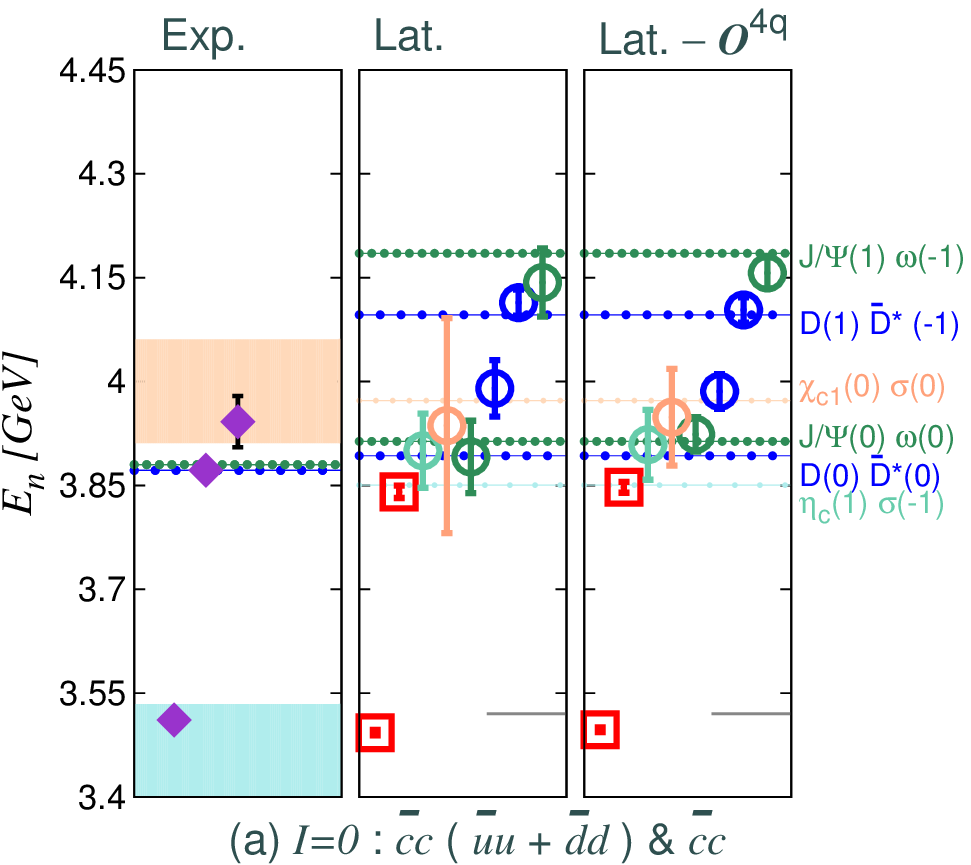}}
\hspace{0.9cm}
\parbox{.45\linewidth}{ 
\centering
\includegraphics[scale=0.72]{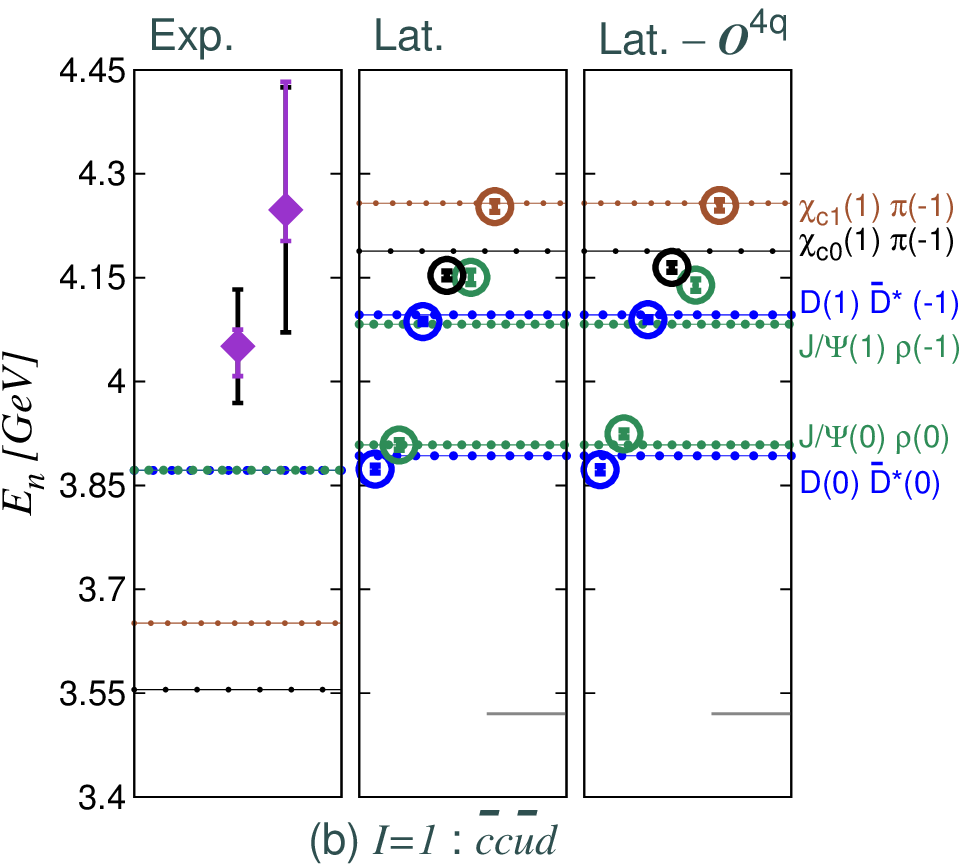}}
\caption{The spectra of states with $J^{PC}=1^{++}$ for the cases with $u/d$ valence quarks. The energies 
$E_n=E^{lat}_n-m_{s.a.}^{lat}+m_{s.a.}^{exp}$ [\eqn{en}] are shown.  The horizontal lines show energies 
of noninteracting two-particle states (\ref{eni}) and experimental thresholds, indicating uncertainty 
related to $\sigma$ width.  In each subplot, the middle block shows the discrete spectrum 
determined from our lattice simulation from the optimized basis [\eqn{optmbasis}].  The right-hand block shows the 
spectrum we obtained from the optimized basis of operators with the $\dad{c}{q}{c}{q}$ operators excluded. 
The gray marks, on the right-hand side of each pane, indicate the lowest three-meson threshold $m_{\eta_c}+2m_\pi $, 
while the actual lowest $\eta_c\pi\pi$ level on the lattice appears higher due to $l=1$, which requires relative momenta. 
The left-hand block shows the physical thresholds and possible experimental candidates (a) $\chi_{c1}$, 
$X(3872)$ and $X(3940)$, (b) $Z_c^+$(4050) and $Z_c^+$(4250). The violet error bars for experimental candidates 
show the uncertainties in the energy and the black error bars show its width. }
\label{fg:primresl}
\end{figure*}

\begin{figure}[tbp]
\centering
\includegraphics[scale=0.72]{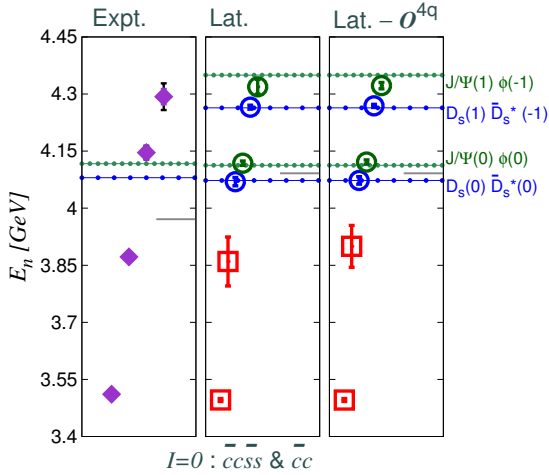}
\caption{The spectrum of states with $J^{PC}=1^{++}$ and hidden strange quarks. The possible 
experimental candidates shown are $\chi_{c1}$, $X(3872)$, $Y(4140)$ and $Y(4274)$.
The gray marks, on the right-hand side of each pane, indicate the lowest three-meson threshold $m_{\eta_c}+2m_K $. 
However, the actual lowest $\eta_cKK$ level on the lattice appears higher due to $l=1$, which requires relative momenta.  
For further details see \fgn{primresl}.}
\label{fg:primress}
\end{figure}

\begin{figure}[tbp]
\centering
\includegraphics[scale=0.72]{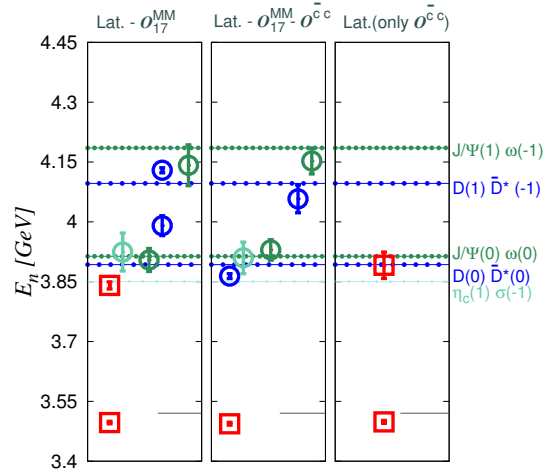}\\
\caption{The spectrum of states (\eqn{en}) with $J^{PC}=1^{++}$ and quark content $\bar cc(\bar uu + \bar dd)$ \& $\bar cc$.  
(i) Optimized basis (without $O^{MM}_{17}$), (ii) optimized basis without  $\bar cc$ operators (and without $O^{MM}_{17}$)
and (iii) basis with only $\bar cc$ operators. Note that candidate for $X(3872)$ disappears when removing $\bar cc$ operators 
although diquark-antidiquark operators are present in the basis, while it is not clear to infer on the dominant nature of this 
state just from the third panel. The $O_{17}^{MM}=\chi_{c1}(0)\sigma(0)$ is excluded from the basis to achieve better signals and 
clear comparison. }
\label{fg:i0sasa}
\end{figure}

\begin{figure*}[tbhp]
\begin{center}
\includegraphics[scale=0.6,clip]{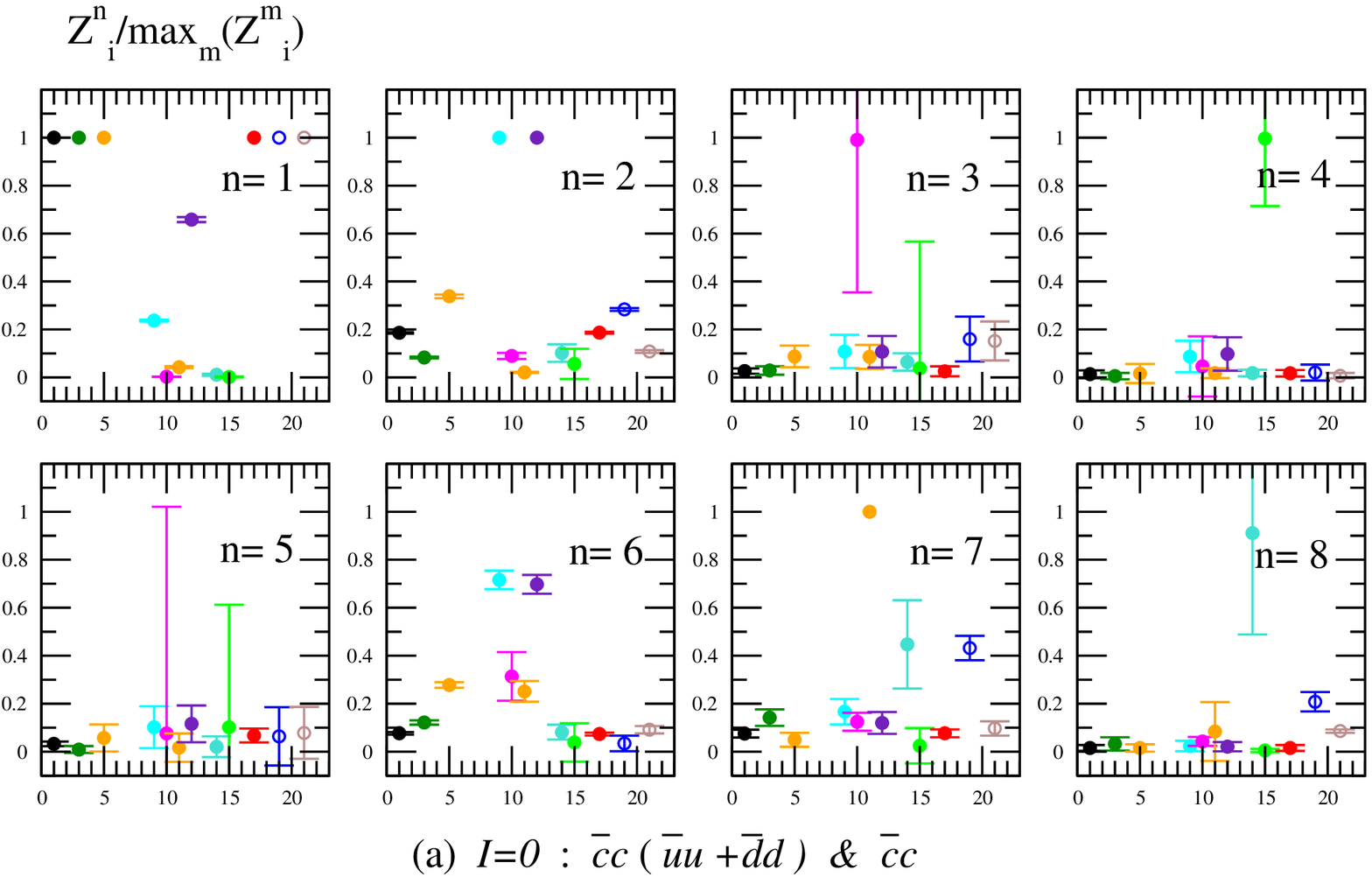}
\end{center}
\begin{center}
\includegraphics[scale=0.6,clip]{I1_w4q.eps}
\end{center}
\begin{center}
\includegraphics[scale=0.6,clip]{ss_w4q.eps}
\end{center}
\caption{The overlap factors $Z_j^{(n)}=\langle \Omega|O_j|n\rangle$  [\eqn{gevp2}] shown in units of the maximal $|Z_j^m|$ for a given operator $j$ across all the eigenstates $m$. 
These ratios are independent of the normalization of the interpolators $O_j$. 
The horizontal axis corresponds to the complete basis of interpolators [Eq. (\ref{operators0})], where the optimized subsets  [\eqn{optmbasis}] were employed. The levels are ordered from lowest to highest $E_n$  as in the middle pane of the spectrum in Figs. \ref{fg:primresl} and \ref{fg:primress}. The values are averages of the ratios over $4 \leq t \leq 13$ with
error bars due to jackknife sampling.}
\label{fg:zratios}
\end{figure*}

\section{Results\label{sec:Results}}

The discrete spectra in Figs. \ref{fg:primresl} and \ref{fg:primress} are the main results from our lattice calculation. They show the energies 
\beq
E_n=E^{lat}_n-m_{s.a.}^{lat}+m_{s.a.}^{exp},\quad m_{s.a.} = \frac{1}{4}(m_{\eta_c}+3m_{J/\psi})~
\eeq{en}
of the states with $J^{PC}=1^{++}$ and three quark contents. The horizontal lines represent various two-meson noninteracting energies. 

The states that have dominant overlap with two-meson scattering operators are represented by circles  and the color coding 
identifies the respective scattering channels  based on the following criteria: 
\begin{itemize}
 \item The levels appear close to the expected two-meson noninteracting energies. 
 \item They have dominant overlaps $\langle\Omega|O_j^{M_1M_2}|n\rangle$ with  corresponding $O_j^{M_1M_2}$. This is also  verified based on the ratios $Z_j^n/max_m(Z_j^m)$,which are independent of normalization of operators and are shown in  \fgn{zratios}.
       
 \item If the corresponding two-meson interpolators are excluded from the basis, this eigenstate disappears or becomes too 
noisy to be identified.  This is determined by comparing the pattern of the effective masses and overlaps between the original basis and the basis after operator exclusion.   
\end{itemize}
The  remaining  states, that are not attributed to the two-meson scattering channels, are represented by red squares. 

Figures \ref{fg:primresl} and \ref{fg:primress} also compare the spectra between the two bases of operators, one with optimized operator set and another with the optimized set excluding $\dad{c}{q}{c}{q}$. In all three cases we see an almost negligible effect on the low lying states,  while we do observe an improvement in the signals for higher lying states in the basis without $\dad{c}{q}{c}{q}$. The same conclusion applies for overlaps. 

The employed irreducible representation $T_1^{++}$ contains the states $J^{PC}=1^{++}$ of interest, 
as well as $J^{PC}=3^{++}$ states due to the broken rotational symmetry. Upon inclusion of the 
interpolator $O_{8}^{\bar cc}$ to the basis [\eqn{optmbasis}] the spectra for both $I=0$ channels 
remain essentially unchanged except for an additional level  at $E\simeq 4.1-4.2~$GeV [\eqn{en}]. This is 
where the earlier simulation on the same ensemble \cite{Mohler:2012na} and the simulation 
\cite{Liu:2012ze} have identified the only $3^{++}$ state in the energy region of our interest. 
In the following subsections, we present the spectra of $J^{PC}=1^{++}$ states in three flavor 
channels for the basis  (\eqn{optmbasis}), where $O_{8}^{\bar cc}$ is excluded. 

\subsection{$I=0$ channel with  flavor $\bar cc (\bar uu+\bar dd)$ and $\bar cc$ \label{ss:I0l}}

This is the channel where the experimental $X(3872)$ resides.  We will argue that the energy levels affected by 
this state are $n=2$ (red squares) and $n=6$ (blue circle) from \fgn{primresl}(a). The lowest state is the 
conventional $\chi_{c1}(1P)$. The overlaps of the three low-lying levels represented by circles show dominant 
$J/\psi(0)\omega(0)$, $\eta_c(1)\sigma(-1)$ and $\chi_{c1}(0)\sigma(0)$ Fock components. The 
highest two states in \fgn{primresl}(a) have significant overlap with the $J/\psi(1)\omega(-1)$ and 
$D_0(1) \bar D_0^*(-1)$ operators. 

Now we focus on the eigenstates that are related to $X(3872)$.  The $\bar cc$ interpolators alone  give 
an eigenstate close to $D\bar D^*$ threshold (right pane of \fgn{i0sasa}), but one cannot establish 
whether this eigenstate is related to $X(3872)$ or to nearby two-meson states in this case. Therefore 
we turn to the spectrum of the full optimized basis [midpane in \fgn{primresl}(a)], where levels 
$n\!=\!2$ (red squares) and $n\!=\!6$ (blue circles) are found to have dominant overlap with the 
$\bar{c}c$ and $D\bar D^*$ operators. Excluding either of these operators results in disappearance of 
one level and a shift in the other level towards the $D\bar D^*$ threshold. We emphasize that one of the two 
levels remains absent when $D\bar D^*$ and $O^{4q}$ are used and $O^{\bar cc}$ is not, as is 
evident from the first and second panel from the left of \fgn{i0sasa}. This indicates that the $\bar cc$ 
Fock component is crucial for $X(3872)$, while the $\dad{c}{q}{c}{q}$ structure alone does not render 
it. This also implies a combined dominance of $\bar{c}c$ and $D\bar D^*$ operators in determining 
the position of these two levels, while their resulting energies are not significantly affected 
whether $O^{4q}$ is used in addition or not.

\begin{table}[htb]
\betb{ c c | c  c  c | c c  }
\hline
$X(3872)$ &&& $m_{X} - m_{s.a.}$ &&& $m_{X} - m_{D_0} - m_{D^*_0}$ \\ \hline
Lat.  &&& 816(15) &&& -8(15) \\
Lat. - $O^{4q}$  &&& 815(8) &&& -9(8) \\
LQCD \cite{Prelovsek:2013cra} &&& 815(7) &&& -11(7) \\
LQCD \cite{Lee:2014uta} &&& - &&& -13(6) \\ \hline
Exp. &&& 803(1) &&& -0.11(21) \\ 
\hline
\eetb
\caption{Mass of $X(3872)$ with respect to $m_{s.a.}$ and the $D_0\bar D_0^*$ threshold. Our estimates are from the 
correlated fits to the corresponding eigenvalues  using single exponential fit form with and without diquark-antidiquark 
operators. Results from previous lattice QCD simulations \cite{Prelovsek:2013cra,Lee:2014uta} and experiment are also presented.}
\label{tb:X3872psa}
\end{table}

We determine the $D\bar D^*$ scattering phase shift from levels $n=2,6$ via L\"uscher's relation 
\cite{Luscher:1990ux} assuming elastic scattering. The phase shift is interpolated near threshold 
using the effective-range approximation. The eigenstate $n\!=\!6$ (blue circle) is interpreted as the 
$D(0)\bar D^*(0)$ scattering state, which is significantly shifted up due to a large negative 
scattering length \cite{Sasaki:2006jn}. The resulting scattering matrix $T\propto 1/(\cot\delta(p) -i)$ has a pole just 
below the threshold where $\cot \delta(p_B)=i$ is satisfied. We neglect possible effects of the 
left-hand cut in the partial wave amplitude.  The results confirm a shallow bound 
state just below the $D\bar D^*$ threshold and the binding momentum $p_B$ renders the mass of the bound 
state, interpreted as experimentally observed $X(3872)$. The resulting mass of $X(3872)$ and its 
binding energy  are provided in \tbn{X3872psa} and in \fgn{X3872psa}, which indicate that it is 
insensitive to inclusion of diquark-antidiquark interpolators within errors. The mass of $X(3872)$ 
was extracted along these lines for the first time in Ref. \cite{Prelovsek:2013cra}, where this 
channel was studied in a smaller energy range on the same ensemble without 
diquark-antidiquark 
interpolators. The error on the binding energy in the present paper is larger due to the larger 
interpolator basis. These results are in agreement with a possible interpretation of X(3872), where its properties are due to the accidental alignment of a $\bar cc$ state with the $D^0\bar D^{*0}$ threshold \cite{Danilkin:2010cc,Takizawa:2012hy}, but we cannot rule out other options.

\begin{figure}[htb]
\includegraphics[scale=0.8]{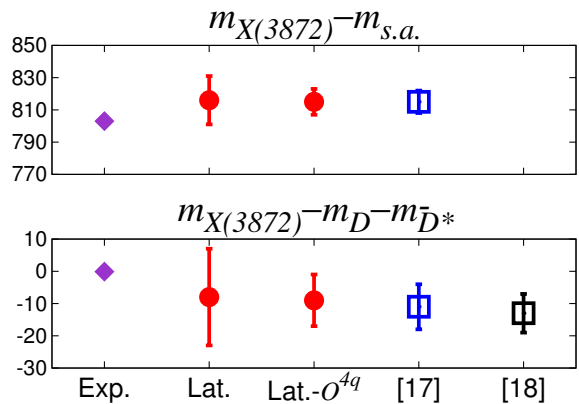}
\caption{Mass of $X(3872)$ with respect to $m_{s.a.}$ from the present simulation, previous lattice studies 
\cite{Prelovsek:2013cra,Lee:2014uta} and experiment \cite{Agashe:2014kda}. }
\label{fg:X3872psa}
\end{figure}
 
With regard to the other experimentally observed charmonia-like states [e.g. $X(3940)$], which could appear in 
this channel, we do not find any candidate in addition to the expected two-meson scattering levels. We also do not 
find candidates for other $\bar cc$ states with $J^{PC}=1^{++}$ [e.g. $\chi_{1c}(nP)$] in the region between 
the $D\bar D^*$ threshold and $4.2~$GeV.

\subsection{$I=1$ channel with flavor  $\bar cc \bar du$ \label{i1}}

A careful analysis of this isospin channel is crucial due to the large branching ratio for the decay 
$X(3872)\rightarrow J/\psi \rho$ and current experimental interests in search of a charged $X(3872)$. 
With no disconnected diagrams allowed in the light quark propagation, the correlation matrix is constructed 
purely of four-quark interpolators and connected Wick contractions in \fgn{Wickcont}(a). 

The spectrum of eigenstates is shown in \fgn{primresl}(b), where all energies are close to noninteracting energy 
levels. All the eigenstates have a dominant overlap with the two-meson interpolators. The spectrum shows very 
little influence on the inclusion of $\dad{c}{q}{c}{q}$, which is evident from \fgn{primresl}(b). Given that 
all the levels below 4.2 GeV can be attributed to the expected two-meson scattering states, we conclude that 
our lattice simulation gives no evidence for $Z_c(4050)^+$ and $Z_c(4250)^+$.

Our results also do not support charged or neutral $X(3872)$ with $I=1$. There is no experimental indication 
for charged $X$, while the neutral $X$ does have a large decay rate to $I=1$ final state $J/\psi \rho^0$. One popular 
phenomenological explanation for this decay is that $X(3872)$ has $I=0$ and the isospin is broken in the decay 
mechanism (due to the $D^+\bar D^{*-}$ vs $D^0\bar D^{*0}$ mass difference) \cite{Gamermann:2009fv,Takizawa:2012hy}. According to another explanation, 
$X$ is a linear combination of $I=0$ and $I=1$ components, where the $I=1$ component vanishes in the isospin limit 
\cite{Tornqvist:2004qy}. Our simulation is performed in the isospin limit $m_u=m_d$, so it 
is perhaps not surprising that $X$ with $I\!=\!1$ is not observed. Future simulations with nondegenerate $u/d$ quarks 
would be very welcome for this channel. 

As pointed out in \scn{Interpolators}, $\rho$ in $J/\psi \rho$ is treated as stable, although $\rho(1)$ is kinematically 
close to the decay channel $\pi(1)\pi(0)$.  In the absence of a simulation of a three-meson system, it is disputable what `noninteracting' energy should be taken for the $\rho(1)$. An estimate 
from the diagonal correlator $\rho(1)$ leads to `noninteracting' energy roughly 65 MeV below the eigenstate energy, 
which is identified to have a dominant overlap with the $J/\psi(1) \rho(-1)$ interpolator. However, taking the resonance 
position \cite{Lang:2011mn} brings the `noninteracting' level in agreement with the measured eigenenergy.

\subsection{$I=0$ channel with flavor  $\bar cc \bar ss$ and $\bar cc$\label{i0s}}
 
Our goal in simulating this channel is to search for a possible presence of the $Y(4140)$ resonance, which was found in 
$J/\psi \phi$ scattering in several experiments \cite{Aaltonen:2009tz}.  
Our lattice simulation of $J/\psi\phi$ scattering takes into account the annihilation of the valence 
strange quarks and thereby the mixing with $\bar cc$ flavor content.

With no strange quark effects in the sea, the study of this channel is based on the following 
assumptions. We construct a basis with only $\bar{c}c$ and four-quark operators ($O^{MM}$, $O^{4q}$) 
with valence hidden strange content for this analysis. We assume that these interpolators have negligible coupling to   two-meson states  with flavor content 
$\bar cc(\bar uu+\bar dd)$. In other words, we assume that two-meson states like $D\bar D^*$ and 
$J/\psi \omega$ will not appear in the spectrum based on the chosen interpolators. 
The resulting spectrum in this channel confirms this assumption. We point out that $Y(4140)$ has been experimentally observed only in the $J/\psi \phi$ final state with valence strange content, but it has not been observed in $D\bar D^*$ and $J/\psi \omega$ final states.  Although this ensemble does 
not have strange quarks in the sea, we assume that the valence strange content could uncover hints 
on the existence of the charm-strange exotics, if they exists. 

Spectra in this channel are shown in \fgn{primress}. We identify the lowest two states, represented 
by squares, to be $\chi_{c1}(1P)$ and the level related to $X(3872)$. The remaining four states 
are identified  with the expected $D_s\bar D_s^*$ and $J/\psi \phi$ scattering levels. Thus in the 
energy region below 4.2 GeV, we find no levels that could be related to $Y(4140)$ or any other exotic 
structure. Note that the existence of  $Y(4140)$ is not yet finally settled from experiment, and  
its quantum numbers, except for $C=+1$, are unknown. Therefore it is possible that its absence in 
our simulation is related to the fact that we explored the channel $J^P=1^+$ only.

\subsection{Discussion}

The only exotic charmonium-like state found in our simulation is a $X(3872)$ candidate with $J^{PC}=1^{++}$ and $I\!=\!0$. 
It is found as a bound state slightly below $D\bar D^*$ threshold and has a mass  close to the experimental mass of $X(3872)$. 
 We point out that this mass corresponds to our $m_\pi\simeq 266~$MeV and was obtained from a rather small lattice volume, 
while chiral and continuum extrapolations have not been performed. Precision determination of its mass with respect to 
$D\bar D^*$ threshold will be a challenging task for future lattice simulation on larger volumes, which also should account
for its coupling with multiple open scattering channels involving two or more hadrons. Recent analytic studies 
consider the quark mass dependence, the volume dependence and the effect from the isospin breaking relevant for future 
lattice studies of $X(3872)$ \cite{Jansen:2013cba}.

Candidates for no other ``exotic" charmonium-like states [except for $X(3872)$] are found in our exploration of the three $J^{PC}=1^{++}$ channels. 
We list several possible reasons for the absence of the energy levels related to other possible exotic states in our simulation:
 
\begin{itemize}

\item  The existence of $Y(4140)$, $Z_c^+$(4050), $Z_c^+$(4250) or any other exotic state in these 
  channels, is not yet settled experimentally. Even if they exist, only $C\!=\!+1$ is established 
  experimentally, while their $J^P$ is unknown. This could explain their absence in our simulation, 
  which probes only $J^P=1^+$.   

\item  Based on the experience, discussed in \scn{methodology}, we expect an additional energy level 
  if the exotic state is a resonance associated to a pole near the real axis in the unphysical Riemann 
  sheet. The absence of an additional energy level could also indicate a different origin of the 
  experimental peak, e.g., a coupled-channel threshold effect.  Further analytical work and 
  lattice simulations are needed to settle the question whether an additional energy level is 
  expected in this case. 

\item Finally, we cannot exclude the possibility that some exotic candidates could be absent due 
  to the relatively heavy pion mass $m_\pi\simeq 266~$MeV, isospin limit $m_u=m_d$, neglect of the 
  charm annihilation contributions, or the absence of the strange dynamical quarks in our simulation.  
\end{itemize}

\section{Conclusions\label{sec:Conc}}

We present the spectra from a lattice QCD simulation of $J^{PC}=1^{++}$ channels with three different 
quark contents: $\bar cc \bar du$, $\bar cc(\bar uu+\bar dd)$ and $\bar cc \bar ss$, where the later 
two can mix with $\bar cc$. The pion mass in this study with $u/d$ dynamical quarks is $m_\pi\!\simeq\! 266~$MeV.  
Using a large number of interpolating fields $\dadt{c}{q}{c}{q}$, $\dads{c}{q}{c}{q}$, 
$(\bar cq)_{1_c}(\bar qc)_{1_c}$, $(\bar cc)_{1_c}(\bar qq)_{1_c}$ and $(\bar cc)_{1_c}$, we extract the spectra 
up to $4.2~$GeV. We find evidence for $\chi_{c1}$ and $X(3872)$, while all the remaining eigenstates 
are related to the expected two-meson scattering channels, which inevitably appear in the dynamical 
QCD. The $\bar cc$ Fock component in $X(3872)$ appears to be more important than the $\dad{c}{q}{c}{q}$, 
since we find a candidate for $X(3872)$ only when $\bar cc$ interpolating fields are used. The $D\bar D^*$ 
interpolators show a more prominent effect on the position of $X(3872)$ than the $\dad{c}{q}{c}{q}$. 
Candidates for charged or neutral $X(3872)$ with $I=1$ are not found in our simulation with 
$m_u\!=\!m_d$, and future simulations with broken isospin would be welcome for this channel.  
We also do not find a candidate for $Y(4140)$ or any other exotic charmonium-like 
structure. Our search for the exotic states assumes an appearance of an additional energy eigenstate on the lattice, 
which is a typical manifestation for conventional hadrons.  Further analytic work is needed to establish whether this working 
assumption applies also for several coupled channels and all exotic structures of interest.

\acknowledgments\label{sec:Ackn}
We thank Anna Hasenfratz and the PACS-CS for providing the gauge configurations. We acknowledge the discussions 
with R. Briceno, L. Leskovec, D. Mohler, S. Ozaki, S. Sasaki and C. DeTar. The calculations were performed on computing 
clusters at  the University of Graz (NAWI Graz), at the Vienna Scientific Cluster (VSC) and at Jozef Stefan 
Institute. This work is supported in part by the Austrian Science Fund FWF:I1313-N27 and by the Slovenian 
Research Agency ARRS Project No. N1-0020. S.P. acknowledges support from U.S. Department of Energy Contract No. 
DE-AC05-06OR23177, under which Jefferson Science Associates, LLC, manages and operates Jefferson Laboratory.

\appendix
\section{Fierz transformation of diquark-antidiquark operators \label{ap:Fierz}}

In this appendix we express the local diquark-antidiquark interpolator as  	
\begin{equation}
O^{4q}(x)=\sum F_i ~M^i_1(x) ~M^i_2(x)
\end{equation}
using Fierz transformations \cite{Nieves:2003in}, where \beq M(x)=\bar q_{a}\Gamma q_a^\prime~(x) \eeq{meson} are local color-singlet currents.  
The momentum projected interpolator $O^{4q}(p)$ is then given by 
\beqa
O^{4q}(p) &=& \sum_i F_i \sum_x e^{ipx} M_1^i(x) M^i_2(x) \nonumber\\ 
& =&   \sum_i \frac{F_i}{V^2} \sum_x e^{ipx} \sum_q e^{-iqx} M_1^i(q) \sum_k e^{-ikx} M^i_2(k)  \nonumber\\
& =&   \sum_i \frac{F_i}{V} \sum_q M_1^i(q)  M^i_2(p-q) 
\nonumber .
\eeqa{fqmp}
Thus the projection to total momentum zero $O^{4q}(p=0)$ can be rewritten as sum over 
two-meson operators with back-to-back momenta. 

Fierz transformation is an operation of rearranging the Fermion fields in a Fermion quadrilinear.
Expressing our local diquark-antidiquark interpolator with explicit color (lower) indices 
and Dirac (upper) indices, we have

\begin{eqnarray}
&&[\bar c ~P ~\bar q]_{\mathcal{G}}[c~N ~q]_{\mathcal{G}}\vert_{{\binom{3_c}{\bar{6}_c}}} =\mathcal G_{abc}\mathcal G_{ade}~\bar c_b^\alpha P^{\alpha\beta}\bar q_c^\beta~c_d^\eta N^{\eta\delta}q_e^\delta\nonumber\\
&&=(\delta_{bd}\delta _{ce}\mp\delta_{be}\delta_{cd}) ~P^{\alpha\beta}N^{\eta\delta}~\bar c_b^\alpha \bar q_c^\beta~c_d^\eta q_e^\delta\nonumber\\
&&= P^{\alpha\beta}N^{\eta\delta}~\bigl\{- (\bar c^\alpha c^\eta)_{1_c} (\bar q^\beta q^\delta)_{1_c} \mp (\bar c^\alpha q^\delta)_{1_c} (\bar q^\beta c^\eta)_{1_c}  \bigr\}\nonumber\\
&&= - (\bar c^\alpha \Gamma_I^{\alpha \eta}c^\eta)_{1_c} (\bar q^\beta G_I^{\beta\delta} q^\delta)_{1_c} \mp (\bar c^\alpha \Gamma_I^{\alpha\delta}q^\delta)_{1_c} (\bar q^\beta H_I^{\beta\eta}c^\eta)_{1_c}  \nonumber\\
&&= - (\bar c ~\Gamma_I ~c)_{1_c} (\bar q ~G_I ~q)_{1_c} \mp (\bar c ~\Gamma_I ~q)_{1_c} (\bar q ~H_I ~c)_{1_c}  
\end{eqnarray}
where we have accounted for $\mathcal G_{abc}\mathcal G_{ade}\vert_{{\binom{3_c}{\bar{6}_c}}}=\delta_{bd}\delta _{ce}\mp\delta_{be}\delta_{cd}$ 
in the second line and a minus sign for Fermion exchange in the third. Each term on the right-hand side 
of the fourth line is expressed as a sum over the index $I=1,...,16$, where $\Gamma_I$ are the elements 
of Clifford algebra $\{\Gamma\}$ and ($G_I$, $H_I$) the unknown coefficient matrices. These coefficient 
matrices can be determined using the orthogonality relation $Tr[\Gamma_I\Gamma_J]=4\delta_{IJ}$ 
\begin{equation}
G_I=\frac14 (N^T\Gamma_I P)^T\quad\mbox{and} \quad H_I=\frac14 (N\Gamma_I P)^T
 \end{equation}
 
The diquark-antidiquark fields can therefore be expressed  as a linear combination of products of two 
color singlet currents with various Dirac structures. Local analogs of our diquark-antidiquark 
interpolating fields can be expressed as 

\begin{widetext}
\beqa
O_{\binom{19}{21}}^{4q} &=& [\bar{c}~C\gamma_5~\bar{u}]_{\mathcal{G}}[c~\gamma_{i}C~u]_{\mathcal{G}} + [\bar{c}~C\gamma_{i}~\bar{u}]_{\mathcal{G}}[c~\gamma_5C~u]_{\mathcal{G}} + K_d \{u\rightarrow d\} \label{opfq1} \\
                                  &=& \mp\frac{(-1)^{i}}2 \{ ~(\bar{c}~\gamma_5~u)(\bar u~\gamma_{i}~c) -~(\bar{c}~\gamma_{i}u)(\bar u~\gamma_5~c) \nn \\
                                   && \quad +~(\bar{c}~\gamma^{\nu}\gamma_5~u)(\bar u~\gamma_{i}\gamma_{\nu}~c) |_{i\ne\nu} ~-~(\bar{c}~\gamma_{i}\gamma_{\nu}~u)(\bar u~\gamma^{\nu}\gamma_5~c) |_{i\ne\nu} \}\nn \\
                                   && +\frac{(-1)^{i}}2\{~(\bar{c}~c)(\bar u~\gamma_{i}\gamma_5~u) +~(\bar{c}~\gamma_{i}\gamma_5~c)(\bar u~u)  \nn \\
                                   && \quad -~(\bar{c}~\gamma^{\nu}c)(\bar u~\gamma_{i}\gamma_{\nu}\gamma_5~u) |_{i\ne\nu} ~-~(\bar{c}~\sigma^{\alpha\beta}~c)(\bar u~\sigma_{\alpha\beta}\gamma_{i}\gamma_5~u)|_{i\ne(\alpha<\beta)}\} \nn \\
                                   && +~ K_d \{u\rightarrow d\} \nn 
\eeqa{opfq12}
\begin{center}
and
\end{center}
\beqa
O_{\binom{20}{22}}^{4q} &=& [\bar{c}~C~\bar{u}]_{\mathcal{G}}[c~\gamma_{i}\gamma_5C~u]_{\mathcal{G}} + [\bar{c}~C\gamma_{i}\gamma_5~\bar{u}]_{\mathcal{G}}[c~C~u]_{\mathcal{G}} + K_d \{u\rightarrow d\}  \\
                                  &=& \mp\frac{(-1)^{i}}2 \{~-(\bar{c}~u)(\bar u~\gamma_{i}\gamma_5~c) +~(\bar{c}~\gamma_{i}\gamma_5~u)(\bar u ~c) \nn \\
                                   && \quad -~ (\bar{c}~\gamma^{\nu}u)(\bar u~\gamma_{i}\gamma_{\nu}\gamma_5~c) |_{i\ne\nu} ~+~(\bar{c}~\sigma^{\alpha\beta}~u)(\bar u~\sigma_{\alpha\beta}\gamma_{i}\gamma_5~c)|_{i\ne(\alpha<\beta)} \} \nn \\
                                   && -\frac{(-1)^{i}}2 \{~(\bar{c}~c)(\bar u~\gamma_{i}\gamma_5~u) -~(\bar{c}~\gamma_{i}\gamma_5~c)(\bar u~u)  \nn \\
                                   && \quad +~(\bar{c}~\gamma^{\nu}c)(\bar u~\gamma_{i}\gamma_{\nu}\gamma_5~u) |_{i\ne\nu} ~-~(\bar{c}~\sigma^{\alpha\beta}~c)(\bar u~\sigma_{\alpha\beta}\gamma_{i}\gamma_5~u)|_{i\ne(\alpha<\beta)}\} \nn \\ 
                                   &&+ ~K_d \{u\rightarrow d\}. \nn
\eeqa{opfqs}
\end{widetext}
 
Various terms resemble two-meson operators $O^{MM}$  [Eq. \ref{operators0}], where $(\bar q \Gamma q^\prime)$ denote color singlet currents.  


\end{document}